\documentclass{optica-article}

\journal{opticajournal} 

\articletype{Research Article}

\usepackage{lineno}

\usepackage{hyperref,amsmath, amsfonts,mathtools, xcolor, ulem, cancel}

\DeclareRobustCommand{\eq}[1]{eq.~\eqref{eq:#1}}

\DeclareRobustCommand{\eqs}[2]{eqs.~\eqref{eq:#1} and \eqref{eq:#2}}

\DeclareRobustCommand{\fig}[1]{Fig.~\ref{fig:#1}}

\DeclareRobustCommand{\app}[1]{appendix~\ref{app:#1}}
\DeclareRobustCommand{\sec}[1]{Sec.~\ref{sec:#1}}

\DeclareRobustCommand{\cH}{\ensuremath{{\cal{H}}}}
\DeclareRobustCommand{\cO}{\ensuremath{{\cal{O}}}}

\DeclareRobustCommand{\change}[1]{{\color{black} #1}}

\begin{document}

\begin{raggedright}
    \hfill MIT-CTP 5593, LA-UR-24-30828\\
\end{raggedright}

\title{Floquet engineering with spatially non-uniform driving fields}

\author{Stella T. Schindler\authormark{1,2,$\dagger$} and Hanan Herzig Sheinfux\authormark{3,*}}

\address{
\authormark{1}Theoretical Division, Los Alamos National Laboratory, Los Alamos, NM 87545, USA
\\
\authormark{2}Center for Theoretical Physics, Massachusetts Institute of Technology, Cambridge, MA 02139, USA
\\
\authormark{3}Physics Department, Bar-Ilan University, Ramat Gan, 5290002 Israel}

\email{\authormark{*}lightmatterinteractions@gmail.com, 
\authormark{$\dagger$}schindler@lanl.gov} 

\begin{abstract*} 
In Floquet engineering, we apply a time-periodic modulation to change the effective behavior of a wave system. In this work, we generalize Floquet engineering to \change{more fully} exploit spatial degrees of freedom, expanding the scope of effective behaviors we can access. We develop a perturbative procedure to engineer space-time dependent driving forces that effectively transform broad classes of tight-binding systems into one another. We demonstrate several applications, including removing disorder, undoing Anderson localization, and enhancing localization to an extreme in spatially modulated waveguides. This procedure straightforwardly extends to other types of physical systems and different Floquet driving field implementations.  
\end{abstract*}

\section{Introduction}
Driving a system periodically can generate new out-of-equilibrium phases that differ substantially from the original undriven system. Such driven phases are ubiquitous and have been extensively studied in physics, from Kapitza's pendulum in mechanics \cite{Kapitza} to Thouless pumping in quantum physics \cite{PhysRevB.27.6083, QNiu_1984}. An important milestone for periodic driving was the discovery of dynamic localization in atomic physics in 1986 \cite{DunlapKenkre}. Naively, one might expect that applying a periodic force to a wavepacket would cause it to accelerate or scatter. However, as the name suggests, in dynamic localization one can drive a lattice containing a single electron in a way that suppresses the coupling between lattice sites, making them behave as if fully disconnected. This counterintuitive influence of dynamic localization has been observed in manifold experiments and is now understood to be a general wave phenomenon \cite{PhysRevLett.67.516,MadisonFischer1998,GRIFONI1998229,PhysRevLett.95.234101, EckardtWeiss2005,Longhi2006,PhysRevLett.99.220403, PhysRevLett.100.190405, EckardtHolthaus2009, Szameit2009,Szameit2010,Struck2012,Struck2013,YuanFan2015}. 

The past two decades have seen a surge of interest in Floquet engineering, which is a considerable extension of dynamic localization and encompasses a range of techniques to modify a wave system's behavior using a periodic driving field.
Experimentally, Floquet driving has been realized in \change{manifold} systems, from condensed matter \cite{PhysRevB.79.081406, PhysRevB.84.235108, Lindner2011, Wang2013, Shan2021, Zhou2023}, to cold atoms \cite{Hofstadter2013,HarperHamiltonian2013, Jotzu2014, PhysRevA.68.013820,PhysRevX.4.031027, Abanin2015, PhysRevA.94.040701, PhysRevX.10.031002}, to photonics \cite{PhysRevLett.83.963,PhysRevB.76.201101,PhysRevLett.103.143903,Garanovich2012,PhysRevB.94.020301,Minkov:16, Sounas2017, Huidobro2019, Yin2022}. Notably, the demonstration of Floquet topological insulators in curved waveguide arrays had a progenitive role in topological photonics \cite{Fang2012,FangYuFan2012,Rudner2013,Rechtsman2013,Tzuang2014,OzawaPrice2016,YuanShiFan2016,Lustig2019}. 
The beauty of Floquet engineering lies in its simplicity: driving a \change{system} periodically gives us new degrees of control, allowing us to generate novel behaviors. At the same time, conventional Floquet engineering is also a rather blunt instrument, with only a \change{few} available parameters (driving strength, frequency, and in some cases polarization \change{and polychromatic waves \cite{PhysRevResearch.4.033213, PhysRevLett.128.066602, PhysRevResearch.5.013123, PhysRevA.107.043309, Neufeld2024, Mitra2024}}) giving us limited leverage to induce a desired behavior in a prefabricated physical platform. \change{So far,} spatial degrees of freedom \change{have been exploited in limited, specific} contexts \cite{Lindner2011,PhysRevLett.110.016802, Morina2018, Sharabi2022, PhysRevLett.128.186802}.

Here, we open up the \change{full set of} spatial degrees of freedom in Floquet engineering, giving us greater control over systems. Specifically, we show that with properly tailored driving fields, we can make a large class of 1D nearest-neighbor tight-binding Hamiltonians behave as any other, to leading order in perturbation theory. This allows us to carry out applications previously thought impossible, such as  undoing Anderson localization \cite{PhysRev.109.1492,PhysRevLett.62.47, Wiersma1997,Schwartz2007, PhysRevLett.98.210401, Hu2008,PhysRevLett.100.013906,doi:10.1063/1.3206091, Martin:11, doi:10.1126/science.1209019, Karbasi2014}; i.e., we show a tailored drive can turn a disordered system into an effectively-ordered one. We also construct a spatially dependent driving that has the opposite effect, inducing Anderson localization in a regular lattice. 

\section{Spacetime-dependent Floquet engineering}\label{sec:general-driving}

Spacetime-dependent Floquet engineering is conceptually straightforward. 
Though our primary physical application in this article is photonic waveguides, it is illustrative to begin with a \change{more} generic example. 
Consider \change{an arbitrary physical} system governed by the time-independent Hamiltonian $\cH$. We wish to drive $\cH$ with a space- and time-dependent field driving $F$ such that the overall system $\cH'$ behaves as a desired time-independent Hamiltonian $\cH_{\rm eff}$. That is,
\begin{align}\label{eq:driven-waveguides}
    \cH'(\vec{x},t) = \cH(\vec{x}) + F(\vec{x},t) \quad \sim \quad \cH_{\rm eff}(\vec{x})
\end{align}
The novel ingredient in this paper is that we allow $F$ to have not just time dependence, but also spatial dependence. We give $F$ a time periodicity $T$ and frequency $\omega = 2\pi/T$. \change{Let us pause to emphasize that this work only explicitly considers \textit{monochromatic} driving fields, whose spatial degrees of freedom will prove quite powerful. Combining this with further degrees of freedom like multiple frequencies on top of spatial dependencies can expand our reach yet further. }

In theory, there may be an infinitely large number of $F$'s we can choose in \eq{driven-waveguides} that adequately approximate the behavior $\cH_{\rm eff}$. 
In practice, however, the range of Hamiltonians $\cH$ that we can fabricate and the fields $F$ we can create may be significantly limited by experimental capabilities. Our goal is to develop a straightforward recipe for engineering driving fields $F$ that approximately produce the desired behavior $\cH_{\rm eff}$, subject to such experimental constraints.

Perturbation theory is one method to approximately solve \eq{driven-waveguides}, as the driving field frequency serves as a natural small expansion parameter. \change{We can obtain } a time-independent approximation of a driven system's behavior \change{using a Magnus expansion \cite{Magnus1954}, a tool originally developed in the applied mathematics literature for solving a class of differential equations that naturally includes many time-periodic systems in physics; as such, it is used extensively in Floquet engineering across many physical platforms, e.g. \cite{doi:10.1080/00018732.2015.1055918, Eckardt2017, Oka2019}.}
Dropping spatial arguments $\vec{x}$ for concision, the Magnus expansion takes the form
\begin{align}\label{eq:magnus}
\int_0^T \!\!\!dt\, \cH_{\rm eff}&=\int_0^t dt_1\, \cH'(t_1) + \frac{1}{2}\int_0^t dt_1 \int_0^{t_1}dt_2\,\left[ \cH'(t_1),\cH'(t_2)\right]\\
	&\null\hspace{-0.32 in}+ \frac{1}{6}\int_0^t\!\!\! dt_1\int_0^{t_1}\!\!\!dt_2 \int_0^{t_2}\!\!\!dt_3 \Big\{\big[\cH'(t_1),\left[\cH'(t_2),\cH'(t_3) \right] \big]+ \big[\cH'(t_3),\left[\cH'(t_2),\cH'(t_1)\right] \big]\Big\}+ ... \,.
 \nonumber
\end{align}
The left-hand side (LHS) of this equation evaluates to simply $T \cH_{\rm eff}$, as $\cH_{\rm eff}$ is time-independent. 
On the right-hand side (RHS), the $n$th order term consists of commutators nested $n$ times. 
The first term on the RHS is $\cO(\omega^{0})$ and describes the dispersion that would occur were $F$ averaged out. The second term on the RHS is an $\cO(\omega^{-1})$ correction. (In \change{dynamic localization (DL)}, these two terms are often treated collectively.) The first term on the RHS may depend on $F$ and differ from the undriven $\cH$. \change{While in principle the Magnus expansion has known convergence issues in various scenarios \cite{PhysRevB.93.144307, doi:10.1080/00018732.2015.1055918, Eckardt_2015, Kuwahara2016}, we will see that in practice, it can serve as a strong starting point for spatially nonuniform Floquet engineering.}

\subsection{\change{Example:} Sinusoidal driving}

Let us consider a simple illustrative example, taking the Hamiltonians \eq{driven-waveguides} to have the form of a matrix, and imposing a sinusoidal driving $F(t) = F_0\sin(\omega t)$, with $F_0$ also a matrix. Evaluating \eq{magnus} over one period of oscillation $T = 2\pi/\omega$, we have that
\begin{align}
    \frac{2\pi}{\omega} \cH_{\rm eff} \approx \frac{2\pi}{\omega} \cH + \frac{2\pi}{\omega^2}[\cH,F_0] - \frac{3\pi}{2\omega^3}[F_0,[\cH,F_0]] + \cO(\omega^{-4})
\end{align}
Finding an appropriate driving field \change{$F(t)$} that makes $\cH_0$ behave as some new effective Hamiltonian $\cH_{\rm eff}$ at $\cO(\omega^{-1})$ accuracy thus amounts to solving an inverse commutator problem:
\begin{align}\label{eq:commutator}
    \cH_{\rm eff} - \cH = \frac{1}{\omega}[\cH,F_0]\,,
\end{align}
where the goal is to calculate $F_0$ in terms of $\cH$ and $\cH_{\rm eff}$. Note that while there are many choices of time-dependent driving fields, they typically reduce to an inverse commutator problem resembling \eq{commutator}. Solving \eq{commutator} for $F_0$ is a straightforward algebraic exercise (see \app{inverse-commutator}).\footnote{Note that \eq{commutator} defines the elements of a Lie algebra. For example, in the case of real $n\times n$ matrices, we have the general linear Lie algebra on the real numbers, $\mathfrak{gl}_n(\mathbb{R})$. Solutions then belong to the commutator subgroup, the special linear Lie algebra $\mathfrak{sl}_n(\mathbb{R})$. Properties of Lie algebras are well known, and restrict the form of what $\cH_{\rm eff}$ in \eq{commutator} we can achieve \textit{exactly}; for example, taking the trace on both sides of \eq{commutator}, we see that $\cH-\cH_{\rm eff}$ must always be traceless. However, we remind the reader that here we are pursuing approximate solutions, and \eq{commutator} is itself an expansion. It is straightforward to consider more general classes of operators as well, such as
non-Hermitian systems \cite{Bender:1998ke, Ruschhaupt_2005, El-Ganainy:07, ChristodoulidesPT2009, Ruter2010}.} 

\section{Floquet engineering of curved waveguides}

As a concrete implementation of spacetime-dependent Floquet engineering, we consider an array of curved paraxial waveguides with low refractive index contrast. In the slowly-varying envelope (paraxial) approximation, waves traveling through this system obey a Schrödinger-like equation
\begin{align}
    i\partial_z \psi(\vec{x},z) = \cH \psi = \Big(\frac{i}{2k_0}\nabla^2 - \frac{k_0}{r_0}[n(\vec{x},z)-n_0]\Big)\psi.
\end{align}
Here, $\cH$ is the Hamiltonian, $\psi$ is the wavefunction, $n(\vec{x},z)$ is the local distribution of the refractive index, and $k_0=2\pi n_0/\lambda$ is the wavenumber. The axis of wave propagation is $z$, which plays the same role as time $t$ in \sec{general-driving}. The vector $\vec{x}$ signifies transverse directions. In the paraxial approximation, the array behaves like a tight-binding model. The curvature of the waveguides induces a $z$-dependent driving field $F(\vec{x},z)$.

\begin{figure}
    \begin{center}
    \raisebox{0.5 in}{\includegraphics[width = 2.8in]{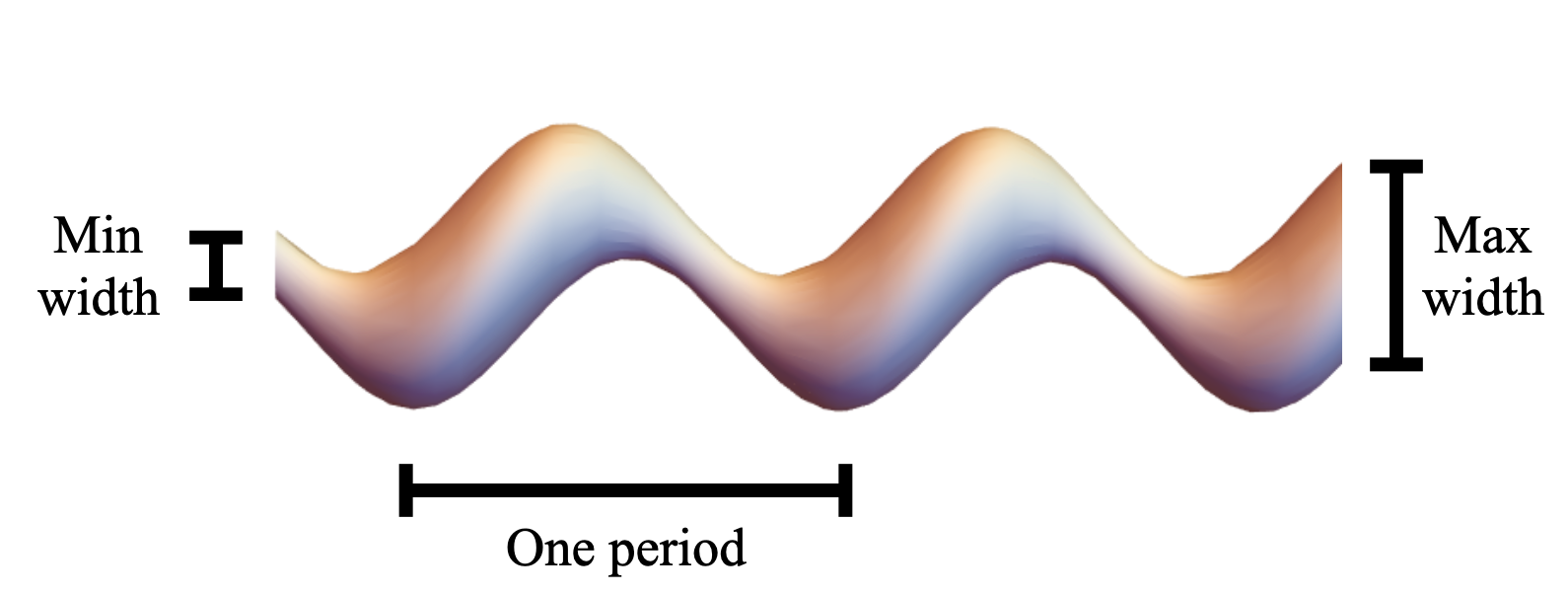}}
    \includegraphics[width = 2.3in]{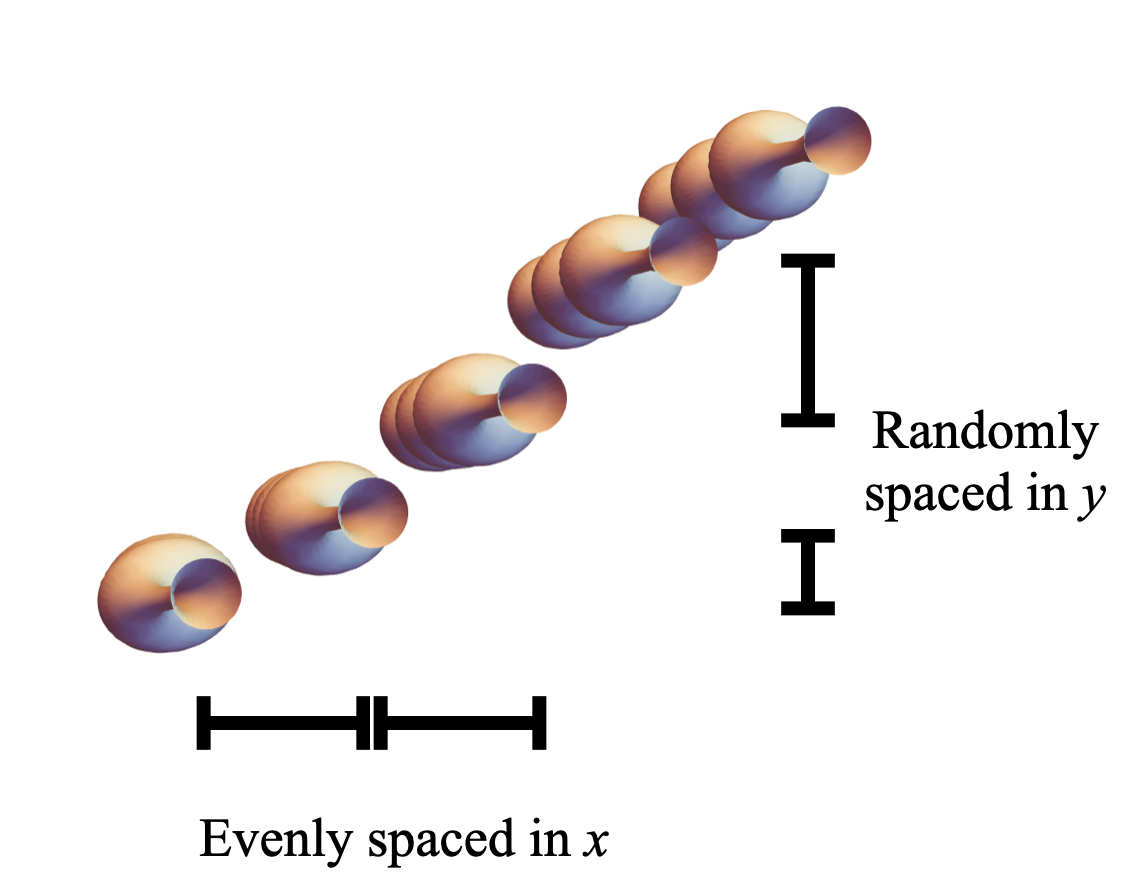}
    \end{center}
    \caption{{\bf System schematic, side and top views.} Set of waveguides evenly spaced in the $x$-direction but randomly spaced along the $y$-axis. Waveguides take periodic trajectories in space and have periodic thickness modulations. Shape is exaggerated for illustration.}\label{fig:system-schematic}
\end{figure}

The simplest such system we can analyze is when the waveguides have only nearest-neighbor (nn) couplings. Straight (undriven) nn-coupled waveguides are described by a $z$-independent tridiagonal matrix
\begin{align}\label{eq:undriven}
    \cH_{\rm 0}=\gamma + \gamma^\dagger \,,
\end{align}
where $\gamma_{ij} = \gamma_i \delta_{i+1,j}$ has nonzero entries only on the first off-diagonal describing the couplings. In the examples used in this paper, we take randomly chosen couplings $\gamma_i \in [0.2,1]$. To induce a driving field in this array, we curve the waveguides along zig-zagged paths, as sketched in \fig{system-schematic}. Concurrently, we undulate the width of each waveguide periodically.
The waveguides oscillate in unison, so their distances are $z$-independent. We make the waveguide trajectories locally parabolic, inducing an artificial gauge field on the wavepackets and causing the nn couplings $\gamma$ to rotate phase as $\gamma_k \exp(\pm i \omega z)$. \change{We vary the on-site potential $V$ and modulate the waveguide thickness as $\epsilon \cos(\omega z)$}, alternating sign $\pm$ after each period. 
The propagation of light through the driven system is governed by the $z$-dependent Hamiltonian 
\begin{align}\label{eq:2}
    \cH_{\rm driven} (z) = \gamma e^{i\omega z} + \gamma^\dagger e^{-i\omega z} + V + \epsilon\cos(\omega z)\,,
\end{align}
as \change{explained in greater detail} in \app{tight-binding} and \app{curved-waveguide-construction}. 

Let us find the lowest-order effective Hamiltonian that describes the behavior of \eq{2}. Applying the Magnus expansion in \eq{magnus} to \eq{2} and averaging over one period,\footnote{Recall that the Schr{\"o}dinger equation carries an extra factor of $-i$, which both lies in $\cH_{\rm eff}$ and propagates throughout the Magnus expansion.} we have
\begin{align}\label{eq:effective-hamiltonian}
    \cH_{\rm eff} = V + \frac{1}{\omega}\Big(2[\gamma,\gamma^\dagger] - [\epsilon,\gamma-\gamma^\dagger] + 2[V,\gamma-\gamma^\dagger] \Big)+ \cO(\omega^{-2})
\end{align}
The terms $V$ and $[\gamma,\gamma^\dagger]$ are both diagonal matrices. The other two commutators  in \eq{effective-hamiltonian} have nonzero elements only on their first off-diagonals. 

Let us now curve the waveguides (i.e., by choosing appropriate $V$ and $\epsilon$) such that $\cH_{\rm eff}$ behaves approximately as some $z$-independent system
\begin{align}\label{eq:desired-hamiltonian}
   \cH_{\rm eff} = V_{\rm eff} + \gamma_{\rm eff} + \gamma^\dagger_{\rm eff}\,,
\end{align}
where $V_{\rm eff}$ is again diagonal and $\gamma_{\rm eff}$ has nonzero entries only on the first off-diagonal. We begin by equating the LHS of \eq{effective-hamiltonian} with the LHS of \eq{desired-hamiltonian}, which simplifies to
\begin{align}\label{eq:matchup}
    [\epsilon,\gamma-\gamma^\dagger]= \Big(\omega V + 2 [\gamma,\gamma^\dagger] -\omega V_{\rm eff} \Big) + \Big(2[V,\gamma-\gamma^\dagger] - \omega\gamma_{\rm eff} - \omega\gamma_{\rm eff}^\dagger
    \Big)
\end{align}
On the RHS, the first parenthetical set of terms is purely diagonal, and the second set of terms is purely off-diagonal. It is straightforward to check that the RHS is purely Hermitian. Likewise, on the LHS, $\gamma-\gamma^\dagger$ is skew-Hermitian. Thus, $\epsilon$ is Hermitian, i.e. $\epsilon_{ij} =\epsilon_{ji}$. 

\begin{figure}[t]
\begin{center}
   \raisebox{0.05 in}{\includegraphics[width = 2.75 in]{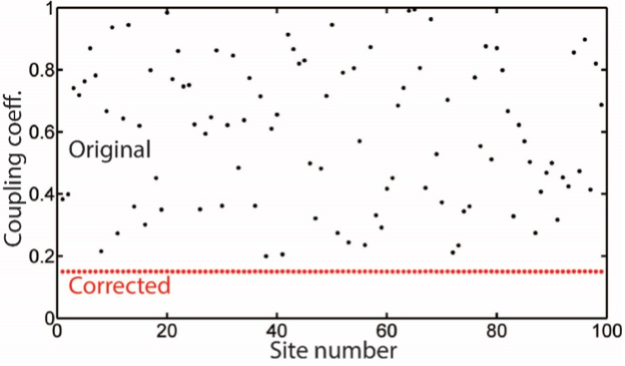}}
    \qquad
   \includegraphics[width = 2 in]{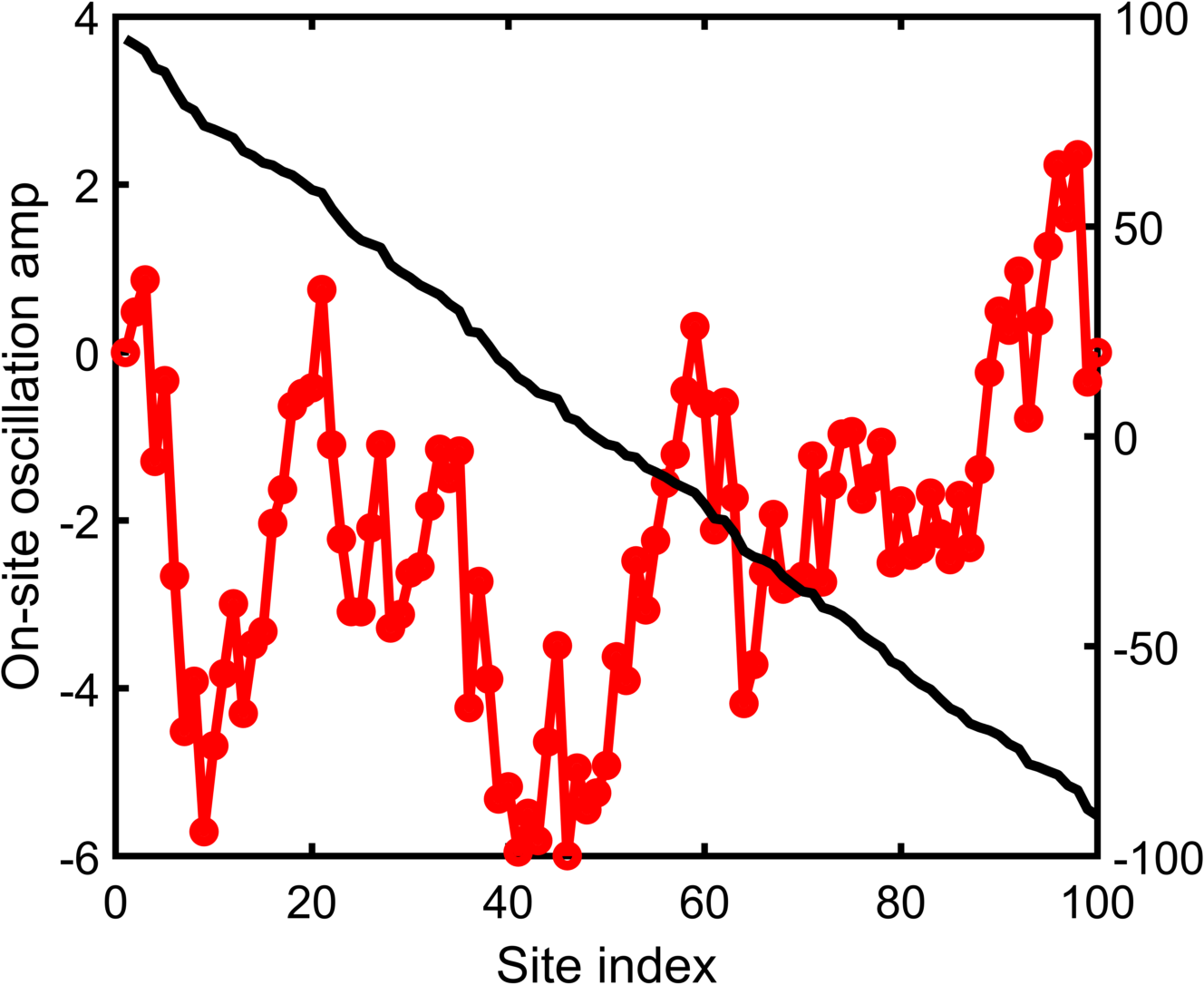}
\end{center}
   \caption{
   \change{
   {\bf Reversing Anderson localization: Couplings.} 
   (Left panel) Black dots represent the couplings in the original undriven system in \eq{undriven}, with random $\gamma_i \in [0.2,1]$. Red dots represent the couplings in the effective Hamiltonian in \eq{effective-hamiltonian}, which are produced when we drive the system as \eq{2} with appropriate parameters $(V,\epsilon)$. 
   (Right panel) To calculate the appropriate driving parameters, we impose $V=0$ and use numerical optimization to calculate $\epsilon_i$, shown here as a black line. \fig{3} verifies that these choices of driving parameters have the expected effect on light propagation through the waveguide array. 
   The significant linear component (scale on right vertical axis) is responsible for uniform changes in the coupling coefficient; this produces a time-dependent electric field, similar to that of dynamic localization. To better exhibit the tailored spatial dependence required to undo Anderson localization, we indicate the smaller nonlinear component of the driving field in red (scale on left vertical axis). 
    }
    }\label{fig:2}
\end{figure}

We have enough degrees of freedom available that we can take $\epsilon$ to be tridiagonal. In this case, let us examine the terms of the $\epsilon$-commutator in \eq{matchup}:
\begin{align}
    [\epsilon,\gamma-\gamma^\dagger]_{ik}
    &= \gamma_{k-1}\epsilon_{i,k-1}  - \gamma_k\epsilon_{i,k+1}- \gamma_i \epsilon_{i+1,k}+\gamma_{i-1} \epsilon_{i-1,k}
\end{align}
This commutator is also tridiagonal. We can examine each diagonal $k = i, i\pm 1$ separately:
\begin{align}\label{eq:comm-diagonals}
    [\epsilon,\gamma-\gamma^\dagger]_{i,i+1} 
    &= \gamma_{i}(\epsilon_{i,i}- \epsilon_{i+1,i+1})\nonumber\\
    [\epsilon,\gamma-\gamma^\dagger]_{ii} 
    &= 2\gamma_{i-1}\epsilon_{i,i-1}\,,
\end{align}
Combining \eq{comm-diagonals} with \eq{matchup}, we find that
\begin{align}
    \epsilon_{i-1,i}=\epsilon_{i,i-1} &= \frac{1}{2\gamma_{i-1}}\Big(\omega V + 2 [\gamma,\gamma^\dagger] -\omega V_{\rm eff}\Big)_{ii}\nonumber\\
    \epsilon_{i,i}- \epsilon_{i+1,i+1} &= \frac{1}{\gamma_i}\Big(2[V,\gamma-\gamma^\dagger] - \omega\gamma_{\rm eff}
    \Big)_{i,i+1}
\end{align}
The latter equation allows us to construct the diagonal terms of $\epsilon$ recursively:
\begin{align}
    \epsilon_{ii} &= \sum_{k=1}^i \frac{1}{\gamma_{n-i}}\Big(2[V,\gamma-\gamma^\dagger] - \omega\gamma_{\rm eff}\Big)_{n-i,n-i+1}\,,
\end{align}
with $\epsilon_{nn} = 0$.  Note that using this scheme, we still have additional degrees of freedom we can exploit; namely, we can alter the on-site potential $V$ to change the structure of $\epsilon$. 

The analytically tailored driving field (curvature) removes disorder up to $\cO(\omega^{-2})$ corrections. If we use large driving frequencies $\omega$, all terms of $\cH_{\rm eff}$ are small, and thus the dynamics are slow and of small bandwidth. In many cases, it is possible to reduce the effect of $\cO(\omega^{-2})$ corrections without explicitly working at next order in the expansion of $H_{\rm eff}$, by numerically optimizing the choice of $V$ and $\epsilon$. 

\subsection{Example driving parameters}\label{sec:example-driving}
Here, let us take our desired Hamiltonian $\cH_{\rm eff}$ to have uniform nearest-neighbor couplings $\gamma^{ij}_{\rm eff} = \Gamma \delta^{i,j+1}$ and no on-site potential $V_{\rm eff} = 0$. 

\paragraph{Example 1.} Let us assume $V=0$. Then our driving field is $\epsilon \cos(\omega z)$, and we can identify:
\begin{align}
    \epsilon_{i-1,i}&=\epsilon_{i,i-1} = \frac{1}{\gamma_{i-1}}(\gamma_i^2-\gamma_{i-1}^2)\nonumber\\
    \epsilon_{ii} &= -\omega \Gamma \sum_{k=1}^i \frac{1}{\gamma_{n-i}}\,.
\end{align}

\paragraph{Example 2.} We could also choose to work with both an on-site potential and a driving field $V + \epsilon \cos(\omega z)$. In this case, it is possible to eliminate the off-diagonal terms in $\epsilon$ by fixing $V$ properly. Specifically, we can take
\begin{align}\label{eq:16}
    V_{ij} &= -\frac{2}{\omega} [\gamma,\gamma^\dagger]_{ij} = -\frac{2}{\omega}(\gamma_i^2 - \gamma_{i-1}^2)\delta_{ij}\nonumber\\
    \epsilon_{i,i\pm 1} &=0\nonumber\\
    \epsilon_{ii} &= \sum_{k=1}^i \frac{1}{\gamma_{n-i}}\Big(-\frac{4}{\omega}\Big[[\gamma,\gamma^\dagger],\gamma-\gamma^\dagger\Big] - \omega\Gamma\Big)_{n-i,n-i+1}
    = -\omega\Gamma \sum_{k=1}^i \frac{1}{\gamma_{n-i}}\,. 
\end{align}

\begin{figure}
    \begin{center}
    \includegraphics[width = 4 in]{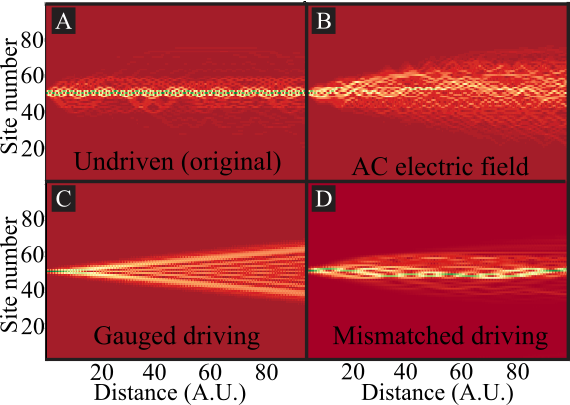}
    \end{center}
    \caption{\change{{\bf Reversing Anderson localization: Light propagation.} Wave propagation in nearest-neighbor coupled waveguides described by \eq{2}, with four choices of driving fields.} (A) An undriven, disordered lattice with random nearest-neighbor couplings. 
    The beam width remains limited, as expected for Anderson localization. (B) 
    \change{The same lattice, driven by a uniform amplitude AC electric field with an amplitude designed to induce dynamic localization for $\omega = 6$; that is, for an $\epsilon_i$ which depends linearly on the site index, such that we would expect dynamic localization if not for disorder.}
    (C) The same lattice, driven by an appropriately tailored time-periodic spatially nonuniform field. The nature of the driving field was determined to make the lattice behave as an effectively disorder-free lattice with uniform couplings, using numerical optimization.
    The beam exhibits discrete diffraction, as expected. 
    \change{We note the wave expansion seen here is indeed extremely similar to the one exhibited in a uniform lattice (not shown here, as it is practically identical), up to the effect of higher order corrections after much longer propagation times.} 
    \change{(D) The same lattice, driven with a random driving field. The wavepacket exhibits diffusive behavior, which is inconsistent with the behavior of both Anderson localization and a regular waveguide array.} }\label{fig:3}
\end{figure}  

\subsection{Reversing Anderson localization} 

In \fig{3}a, we show how waves propagate through a set of straight waveguides, as described by \eq{undriven}. These waveguides are disordered, so we expect Anderson Localization to bring transverse transport in the array to a halt \cite{Segev2013}. In \fig{3}a, we see that a localized beam incident on the array does indeed remain localized. 

In \fig{3}b, we consider propagation through the waveguide when acted upon by a simple, spatially-uniform AC driving field. \change{This case corresponds to an $\epsilon_i$ which is proportional to the site index, so that the difference between every two sites (the electric field) is constant. Whatever the field amplitude, we do not expect the localizing effect of disorder to be undone, so as an example we take the amplitude for which an equivalent ordered lattice would not undergo localization.} As expected, an AC field generally causes a wavepacket to diffuse and eventually arrests its propagation.\footnote{\change{Except in extreme cases, like when the field strength is so large as to completely dominate the Hamiltonian's behavior and dwarf the disorder.}}

In \fig{3}c, we construct one possible curvature of the waveguides (driving field) in \eq{2} that counteracts the disorder in the waveguides, making the system behave as an effectively uniform, ordered array. To do so, we calculate appropriate choices of $V$ and $\epsilon$ that produce the desired effective Hamiltonian. Specifically, we make a choice of $\epsilon$ that is pure diagonal, as described in \sec{example-driving} \change{and \fig{2}}. Note that to construct the driving field used in this figure, we first carry out calculations using the procedure outlined above, which we use as the starting point for a numerical optimization of diagonal matrices $V$ and $\epsilon$ producing the desired behavior.

\change{In \fig{3}d, we show that} in contrast, applying an arbitrary driving force typically results in diffusive behaviors. In some cases, time-dependent disorder can even lead to anomalous faster-than-ballistic diffusive transport \cite{Levi2012}. For our curved waveguides, we also find diffusive behavior if the driving parameters are arbitrarily chosen. This is in sharp contrast with the elimination of disorder that appears when the spatial dependence of the drive is tailored according to our prescription. The elimination of disorder we can obtain by appropriately driving the system is therefore in sharp contrast with typical regime of periodically driven disorder.

We obtain these results for relatively low $\omega=6$ where higher order correction terms cannot be outright neglected. \change{(For a sense of scale, recall that the random couplings $\gamma_i$ are taken to be $\cO(1)$.)} Since the magnitude of $\cH_{\rm eff}$ is proportional to $\omega^{-1}$, the dynamics of the driven system are comparable in speed to the dynamics of the undriven system, which is a clear advantage for experiments. Nevertheless, the influence of $\omega^{-2}$ effects is seen to be practically small, especially if further numerical optimization of the driving parameters is performed.

\subsection{Inducing dynamic localization} 
\begin{figure}
    \begin{center}
    \includegraphics[width = 2 in]{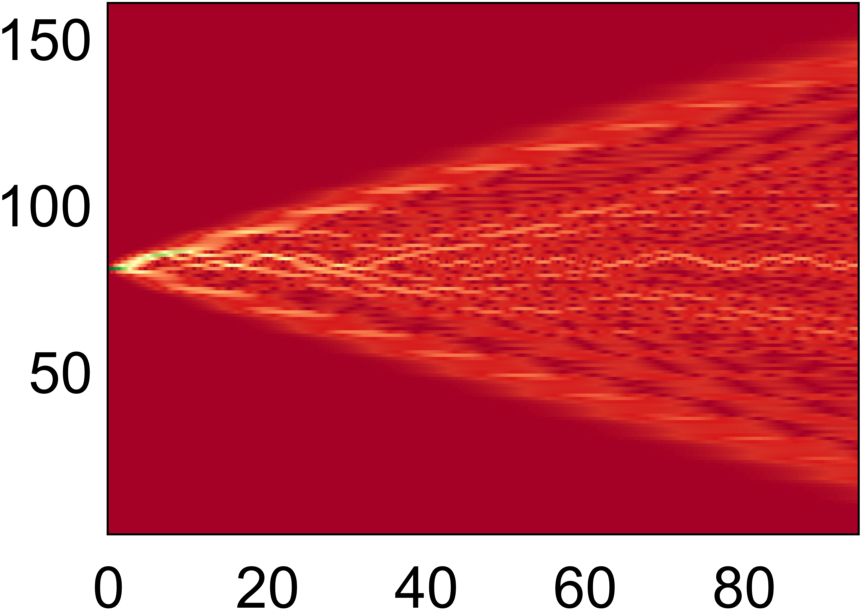}
    \qquad
    \includegraphics[width = 2 in]{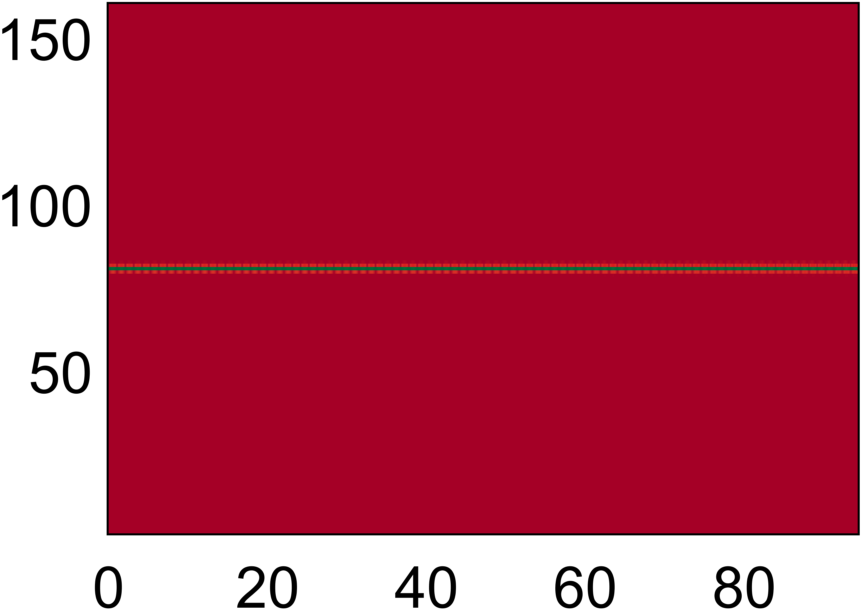}
    \end{center}
    \caption{\change{{\bf Inducing dynamic localization:} (Left panel) This undriven irregular waveguide array has a coupling which changes as a cosine of the index number from 1 to 200, which leads to a complex diffraction of a single site excitation. (Right) Using a tailored spatially non-uniform driving, we obtain notable suppression of the wavefunction spreading. The driving profile is produced in the same method as before; the target Hamiltonian is completely decoupled.}}\label{fig:dl}
\end{figure}

Another capability of our approach is the generalization of dynamic localization \change{(DL), as shown in \fig{dl}.} DL is typically only considered for spatially regular (periodic) lattices; it is commonly believed to fail in inhomogeneous lattices, with the notable exception of driven Glauber-Fock lattices \cite{Longhi2012}. The situation changes remarkably if we use a spatially inhomogeneous driving: it is clear from our explorations of \eqs{magnus}{2} that an appropriately tailored driving field could induce DL even in a lattice which is completely irregular. This gives us the powerful ability to sever specific links and effectively cut out parts of a lattice, by driving it with an appropriate spatially inhomogeneous field. For example, we could create an artificial sharp edge in a topological system like in Refs. \cite{Jotzu2014, Hofstadter2013} and expect to see edge modes there. This technique extends to complicated, highly-nonuniform, and even multidimensional lattices, which all obey a similar operator form to \eq{magnus}.

\section{Generalizations}

Floquet engineering has of course seen extensive applications beyond \change{the} curved waveguides \change{we have examined here. For example, we could study different physical platforms, alternative methods of implementing a driving field, or even tinkering with additional degrees of freedom such as frequencies.}
Spatially nonuniform Floquet engineering techniques can \change{straightforwardly} be applied to these systems as well. 
When \change{constructing a spatially-nonuniform Floquet driving field in any given case, one} must strike a delicate balance between challenges in the mathematical calculation and physical implementation. 

\subsection{\change{Magnus expansion}}
When using perturbative methods to design a driving field, the nature of the Magnus expansion places restrictions on the Hamiltonians and driving fields. 
For example, to solve \eq{magnus} at $\cO(\omega^{-1})$, we generally must solve an inverse commutator problem of the form $\cH_{\rm eff} = [\cH_0,F]$. If $\cH_{\rm eff}$ and $\cH_0$ are both Hermitian, then $F$ must be skew-Hermitian. Likewise,  if $\cH_{\rm eff}$ and $F$ are Hermitian, then $\cH_0$ is skew-Hermitian. Skew-Hermitian waveguide schemes or driving fields may be difficult to fabricate. This does not, however, render Magnus expansion methods impracticable. Indeed, our curved waveguide setup satisfied the skew-Hermitian criterion, and other potential implementations exist.

Alternatively, one could circumvent the problem of skew-Hermiticity by choosing a driving field $F$ such that the $\cO(\omega^{-1})$ term $[\cH_0,F]$ vanishes and the leading term in the Magnus expansion is an $\cO(\omega^{-2})$ integral over the double-commutator $[F,[\cH_0,F]]$. In this case, $\cH_{\rm eff}$, $\cH_0$, and $F$ can be Hermitian simultaneously. However, even if $\cH_{\rm eff}$ and $\cH_0$ are tridiagonal, $F$ may have nonzero terms on its second off-diagonal (or vice versa). This has its own set of challenges for experimental realization.

\subsection{\change{Beyond the Magnus expansion}}

\change{The leading-order Magnus expansion provides a straightforward demonstration of the power of spacetime-dependent Floquet engineering. However, a na{\"i}ve Magnus expansion is not optimal for solving high-precision Floquet engineering problems at low computational cost, and more creative approaches are necessary to full capitalize on the power of spacetime modulations, within physical constraints. For example,
one could consider alternative approximation schemes (see e.g. \cite{RodriguezVega2021}) or simply use numerical optimization to determine a spatially-nonuniform driving field that produces as close to the desired behaviors as possible. The potential for harnessing the full power of spacetime Floquet engineering for physical applications is bright.}

\section{Outlook} 
We open up Floquet engineering to \change{the full breadth of} of spacetime-dependent driving fields, giving us greater control over wave systems. By exploiting spatial degrees of freedom, we showed that we could both induce dynamic localization in non-uniform lattices and undo Anderson localization by giving photonic waveguides appropriate curvature. The technique is more broadly applicable, and we foresee its use in a range of atomic, optical, and condensed matter systems. In the future, perhaps this technique can help in fabricating classes of Hamiltonians that currently are difficult or implausible to realize, such as imaginary gauge fields, Hamiltonians with only next-nearest-neighbor couplings, or non-Abelian Hamiltonians.

\begin{backmatter}

\bmsection{Funding} 
S.T.S. was supported by the U.S. Department of Energy, Office of Science, Office of Nuclear Physics from DE-SC0011090; the U.S. National Science Foundation through a Graduate Research Fellowship under Grant No. 1745302; fellowships from the MIT Physics Department and School of Science; and the Hoffman Distinguished Postdoctoral Fellowship through the LDRD Program of Los Alamos National Laboratory under Project 20240786PRD1. Los Alamos National Laboratory is operated by Triad National Security, LLC, for the National Nuclear Security Administration of the U.S. Department of Energy (Contract Nr. 892332188CNA000001). 

\bmsection{Acknowledgments} 
We thank Mordechai Segev and Yaakov Lumer for initial work on this project many years past \cite{8083203}, and S.T.S. is grateful to Moti for hosting her as a summer research student during that time. We thank Momchil Minkov and Pengning Chao for interesting discussions.

\bmsection{Disclosures} 

\bmsection{Data Availability Statement}

\end{backmatter}

\bibliography{floquet}

\appendix
\section{Inverse commutator problems}\label{app:inverse-commutator}
Consider an inverse commutator problem of the form in \eq{commutator}:
\begin{align}\label{eq:commutator2}
    A = [B,C]\,.
\end{align}
The physical constraint of Hermiticity of $A$ and $B$ restricts the form of $C$:
\begin{align}
    A^\dagger = [B,C]^\dagger = C^\dagger B^\dagger - B^\dagger C^\dagger = C^\dagger B - BC^\dagger = [B,-C^\dagger]\,.
\end{align}
From $A=A^\dagger$ and $A$ real, we need $C$ to be skew-Hermitian, i.e., $C = -C^\dagger$.
Skew-Hermiticity and reality of $C$ implies that its diagonal vanishes:
\begin{align}
    C_{ii} = 0
\end{align} 
for all $i$. 
To solve for components $C_{ij}$ with $i\neq j$, we first label the eigenvalues and
eigenvectors of $B$ as $B\vec{v}_i = E_i \vec{v}_i$.
We arrange the eigenvectors into a matrix $V = (\vec{v}_i,\,...,\,\vec{v}_n)$. Multiplying \eq{commutator2} on both sides by $V$ and $V^{-1} = V^\dagger$ and examining the $(i,j)$, we have that
\begin{align}
    (V^{-1}AV)_{ij} = \vec{v}_i^{\,T}[B,C]\vec{v}_j = (E_i - E_j)(V^{-1}CV)_{ij}\equiv (E_i-E_j)\tilde{C}_{ij}\,.
\end{align}
Thus, for $i \neq j$, we have that
\begin{align}\label{eq:c-solve}
    \tilde{C}_{ij} =(VCV^{-1})_{ij} = \frac{\vec{v}_iA\vec{v}_j^{\,T}}{E_i-E_j} \,.
\end{align}
We also note that this is only consistent with $C$ being real skew-Hermitian if and only if
\begin{align}
    \vec{v}_i^{\,T}A\vec{v}_i \stackrel{?}{=} 0,
\end{align} 
which is in general \underline{not} true. However, in practice, this usually is not a major concern, given that we are working only to lowest order in a perturbative expansion.

Now that we know the entries of $\tilde{C}$ from \eq{c-solve}, it is straightforward to solve for $C$ by simply multiplying on the left by $V^{-1}$ and on the right by $V$:
\begin{align}
    C = V\tilde{C} V^{-1}\,.
\end{align}
We emphasize again that spatially-nonuniform Floquet engineering schemes can be considered for more general classes of operators as well.

\section{Tight-binding model derivation}\label{app:tight-binding}
In this section, we review the tight-binding model equivalent of a waveguide lattice. Next, we generalize it to the case of waveguides that have oscillating widths and different trajectories. To the best of our knowledge, the latter derivation has not explicitly been written down elsewhere in the literature; it may thus be of interest to the reader. 

It is useful to first analyze the case of a single waveguide, which we take as centered about the curve $\vec{r} = [x(z),y(z),z]$ with shape function $u(\vec{r})$. Under the slowly-varying envelope (paraxial) approximation, light propagating through this waveguide is described by the paraxial wave equation
\begin{equation}\label{eq:paraxial}
	i\frac{\partial}{\partial z}\psi(\vec{r}) = -\frac{1}{2k_0}\left( \frac{\partial^2}{\partial x^2} + \frac{\partial^2}{\partial y^2}\right)\psi(\vec{r}) - \frac{k_0}{n_0}u(\vec{r})\psi(\vec{r}).
\end{equation}
This equation is isomorphic to the Schrödinger equation, albeit with the propagation direction $z$ of waves in optics playing the traditional role of time $t$ in quantum mechanics. The wavenumber is $k_0 = 2\pi n_0/\lambda$, where $n_0$ is the refractive index of the bulk material and $\lambda$ is the wavelength of the light. The shape function can be related to the total refractive index as $u(\vec{r}) = n(\vec{r})-n_0$.

We let the waveguide shape evolve slowly and remain localized around $\vec{r}$, so that it is a simple single-mode waveguide. We denote the local phase accumulation speed of the propagating mode as $\beta(z) = \beta_R(z) + i\beta_I(z)$, where $\beta_I$ is a $z$-dependent loss term arising from coupling to continuum modes. If $u(\vec{r})$ changes adiabatically (i.e. slowly relative to $\beta_R$), then this rate of loss is small with $\beta_I \ll \beta_R$. This statement describes many physical systems. For example, an array of low-contrast waveguides created by laser-writing techniques has strongly paraxial waveguides, and $\beta_I$ contributes to a uniform decay of light in all waveguides within the array. We remark here that even small decay can have a major practical effect in certain cases, like long arrays.

Next, let us consider a set of $n$ single-mode waveguides labeled $k = 1,2,...,n$ centered about the curves $\vec{r}_k = [x_k(z), y_k(z),z]$. Here, $z$ is the longitudinal direction along which waves propagate, and $x_k(z)$ and $y_k(z)$ are functions describing how the waveguides curve through space in the $x$-$y$ plane, i.e., transverse to the propagation direction $z$. The width of each of these waveguides varies in space, and we describe the radius of a waveguide in the $x$-$y$ plane at each point along its trajectory in the $z$-direction by the shape function $u_k(\vec{r}) = U(\vec{r_k})\delta(\vec{r}-\vec{r}_k)$. We now can write the behavior of the full array as
\begin{align}
    u(\vec{r}) = \sum_{j=1}^n u_j(\vec{r}) 
\end{align}
In the case discussed in this article, we choose the
waveguides to be equally spaced in the $x$-direction (i.e. $x_1 = x_0$, $x_2 = 2x_0$, $x_3 = 3x_0$, ... ) with the same periodically curved trajectory $X(z)$, i.e.,
\begin{align}\label{eq:setup}
    x_k(z) = k x_0 + X(z)\,.
\end{align}
We distribute the waveguides at random spacings relative to one another along the $y$-direction. 
We write the local mode profile of the wave as $\phi_k(x,y) \equiv \phi\big(x_k(z),y\big)$ and the complex phase accumulation rate at the $k$th waveguide as $\beta_k(z)$. Note that for all $k = 1, ..., n$, we have that
\begin{align}\label{eq:modes}
    \left[-\frac{1}{2k_0} \left(\frac{\partial^2}{\partial x^2} + \frac{\partial^2}{\partial y^2} \right) + \sum_{j \neq k}u_j(\vec{r}) \right]\phi_k = \beta_k \phi_k\,. 
\end{align}
Let us write a wavepacket as the sum
\begin{align}\label{eq:wavepacket}
    \psi(\vec{r}) = \sum_{k=1}^n a_k(z) \phi_k(x,y)\,,
\end{align}
where $a_k(z)$ are a set of coefficients. Plugging \eq{wavepacket} into \eq{paraxial}, we have that
\begin{align}
    i\sum_{k=1}^n \left[\frac{\partial a_k}{\partial z} + a_k\frac{\partial x_k}{\partial z}  \frac{\partial}{\partial x}\right]\phi_k = 
    -\sum_{k=1}^n \left[\frac{1}{2k_0}a_k\left( \frac{\partial^2}{\partial x^2} + \frac{\partial^2}{\partial y^2}\right) + \frac{k_0}{n_0}a_k u(\vec{r})\right]\phi_k
\end{align}
We can rearrange terms to find
\begin{align}
    i\sum_{k=1}^n \frac{\partial a_k}{\partial z}\phi_k &= -\sum_{k=1}^n \left\{\frac{1}{2k_0}\left[
    \left(\frac{\partial}{\partial x} +ik_0 \frac{\partial x_k}{\partial z} \right)^2 +\frac{\partial^2}{\partial y^2}\right]+ \frac{k_0}{2}\left( \frac{\partial x_k}{\partial z} \right)^2  + \frac{k_0}{n_0} u(\vec{r})\right\}a_k\phi_k\,.
\end{align}
Next, we make the transform $\phi_k \to  \exp\big[-ik_0\frac{\partial x_k}{\partial z}\left(x-x_k\right)\big]\phi_k$, giving us
\begin{align}\label{eq:intermediate-1}
    i\sum_{k=1}^n e^{-ik_0\frac{\partial x_k}{\partial z}\left(x-x_k\right)}\frac{\partial a_k}{\partial z}\phi_k &=-\sum_{k=1}^n e^{-ik_0\frac{\partial x_k}{\partial z}\left(x-x_k\right)}\left[\frac{1}{2k_0}\left(\frac{\partial^2}{\partial x^2} +\frac{\partial^2}{\partial y^2}\right)+ \frac{k_0}{2}\left( \frac{\partial x_k}{\partial z} \right)^2  + \frac{k_0}{n_0} u(\vec{r})\right]a_k\phi_k\,.
\end{align}
Then, we plug \eq{modes} into \eq{intermediate-1}, giving us
\begin{align}
    i\sum_{k=1}^n e^{-ik_0\frac{\partial x_k}{\partial z}\left(x-x_k\right)}\frac{\partial a_k}{\partial z}\phi_k
    &= -\sum_{k=1}^n e^{-ik_0\frac{\partial x_k}{\partial z}\left(x-x_k\right)}\left[\beta_k + \frac{ k_0}{2}\left( \frac{\partial x_k}{\partial z} \right)^2  + \left(\frac{k_0}{n_0} -1\right)u(\vec{r})+ u_k(\vec{r})\right]a_k\phi_k
\end{align}
Next, we multiply both sides by $\sum_j \phi_j^* \exp\big[ik_0\frac{\partial x_j}{\partial z}\left(x-x_j\right)\big]$. Integrating over $x$, we get
\begin{align}\label{eq:intermediate-2}
    -i S_{jk} \frac{\partial a_k}{\partial z} = \epsilon_j S_{jk} a_k + \Gamma_{jk}a_k + \Delta_{k}a_k\,,
\end{align}
where we have defined 
\begin{align}
    S_{jk} &= \exp\left[ik_0  \left(\frac{dx_k}{dz}x_k - \frac{dx_j}{dz}x_j \right) \right]\int dx\, \phi_j^* \phi_k \exp\left[ik_0 x \left(\frac{\partial x_k}{\partial z} - \frac{\partial x_j}{\partial z} \right) \right]\nonumber\\
    \epsilon_k &= \beta_k + \frac{k_0}{2}\left( \frac{\partial x_k}{\partial z} \right)^2\nonumber\\
    \Gamma_{jk} &= \left(\frac{k_0}{n_0} -1\right)\exp\left[ik_0  \left(\frac{dx_k}{dz}x_k - \frac{dx_j}{dz}x_j \right) \right]\int dx\, \phi_j^* u(\vec{r}) \phi_k \exp\left[ik_0 x\left(\frac{\partial x_k}{\partial z} - \frac{\partial x_j}{\partial z} \right) \right]\nonumber\\
    \Delta_{k} &= \int dx\, u_k(\vec{r}) |\phi_k|^2\,.
\end{align}
Here, we make the crucial assumption that the modes are tightly-bound matrices and the $S_{jk}$ matrix is purely real and mostly diagonal: $S_{jj}/S_{jk} \approx \delta_{jk}$. 
The $\Gamma_{ij}$ terms are more nuanced and may lead to complicated behavior.


Typically, we can make significant simplifications to \eq{intermediate-2} if $U$ is a narrow potential well that changes on a short distance scale relative to $\left(k_0 \frac{dx}{dz} \right)^{-1}$; using \eq{setup}, we write
\begin{equation}
    \Gamma_{jk} \propto \exp \left[ik_0 (x_k-x_j)\frac{dX}{dz} \right]\int dx\,  \phi_j^* U_k \phi_k\,.
\end{equation}
Using the nearest-neighbor approximation, we have that
\begin{equation}
    \int dx\,  \phi_j^* U_k\phi_k \propto \delta_{j+1,k} + \delta_{j-1,k}\,.
\end{equation}
We thus obtain a standard tight-binding model:
\begin{equation}
    i\frac{da_k}{dz} = \epsilon_k a_k + c_{k,k+1} e^{i\theta_k}a_{k+1} + c_{k,k-1}e^{-i\theta_k}a_{k-1}
\end{equation}
where we have written that $\theta_k = k_0 (x_k-x_j) \frac{dX}{dz}$ and $c_{k,k+1} = c_{k,k-1} = |\Gamma_{jk}|/S_{kk}$.

\section{Constructing curved waveguides}\label{app:curved-waveguide-construction}

We now examine how to physically construct a Hamiltonian like \eq{2} using curved waveguides. Because we are creating a periodic potential, we need waveguides with a shape that is periodic along the $z$-direction. We let the phase accumulation rate change as $\beta_i(z) = \beta_0 + \beta_i \cos(\omega z)$. There are several ways to fabricate such a system. First, we could modulate the depth of the waveguide. This would also change the magnitude of $\beta_i$, but this change is often small relative to $\beta_0$ and thus would have a negligible effect on system dynamics. A second option is to change the shape of the potential asymmetrically so that the area of the waveguide (and thus the value of $\beta$) would change significantly, whereas the coupling coefficient in a particular direction would not. It is most practicable to carry out a full calculation, accounting for and correcting for the effects of a $z$-dependent coupling.

To induce a rotating phase, we oscillate the waveguide trajectories in unison, as described in \eq{setup}. One possible trajectory we could take is $x_k = kx_0 + \frac{\alpha}{2}z^2$, where $x_0$ is the separation between waveguides in the $x$-direction, similar to that of \cite{Longhi2006}.
However, this trajectory rapidly diverges, which is a hindrance to experiment. So, instead we choose the convergent, scalloped trajectory
\begin{align}\label{eq:trajectory}
    x_k(z) = \begin{cases}
        kx_0 + \frac{\alpha}{2}\left(z - \tau \left\lfloor \frac{z}{\tau}\right\rfloor \right)^2 &\qquad  0 < (z\,\,{\rm mod}\,\, 2\tau) < \tau\\
        kx_0 + \frac{\alpha}{2}\left[1-\left(z - \tau \left\lfloor \frac{z}{\tau}\right\rfloor \right)^2\right] &\qquad  \tau < (z\,\,{\rm mod}\,\, 2\tau) < 2\tau
    \end{cases}
\end{align}
Here $\lfloor ... \rfloor$ indicates the floor operation. The scalloping is $2\tau$-periodic and has kinks at every $\tau$; however, mild experimental smoothening should not be detrimental to the working principles behind this Floquet engineering scheme.

Because \eq{trajectory} is locally parabolic,  it gives rise to linear accumulation of phase
\begin{equation}
	\theta_k = k_0 (x_{k+1}-x_k)\frac{dx_k}{dz} = \pm k_0 x_0 \alpha z.
\end{equation}
If we set $\alpha = \omega/(k_0x_0)$, we obtain the desired phase rotation of the coupling coefficient. Recall, however, that the waveguides are not uniformly distributed in the $y$-direction, which results in an overall nonuniform coupling coefficient
\begin{align}
    \gamma_i = {\rm exp} \left[-\kappa \sqrt{x_0^2 - (y_{i+1}-y_i)^2} \right]
\end{align}
with $\kappa$ the modal decay distance. If we take $y_i \in (0,x_0]$, then $\gamma_i\in [e^{-\kappa x_0},1]$.

As a consequence, we have an experimentally-realizable setup that is similar but not identical to \eq{2}:
\begin{align}\label{eq:desired-h-2}
	\cH = \begin{cases}
		\gamma e^{i\omega t} + \gamma^\dagger e^{-i\omega t} + V + \epsilon \cos(\omega t) & \qquad 0 < (z\,\,{\rm mod}\,\, 2\tau) < \tau\\
		 \gamma e^{i\omega t} + \gamma^\dagger e^{-i\omega t} + V - \epsilon \cos(\omega t)& \qquad \tau < (z \,\,{\rm mod}\,\, 2\tau) < 2\tau
	\end{cases}
\end{align}
Here, just as in the main text, we assume that the sign of $\epsilon$ changes every half-cycle $\tau$. This again would create a discontinuity in an experiment.

\end{document}